\documentclass[a4paper,journal,twocolumn,twoside,10pt]{IEEEtran}

\usepackage{amssymb}
\usepackage[cmex10]{amsmath}
\usepackage{units}
\usepackage{graphicx}
\usepackage{color}
\usepackage{mathtools}
\usepackage[noadjust]{cite}

\usepackage{expl3}
\makeatletter
\ExplSyntaxOn
\cs_gset_eq:NN \c_deprecateacro_minus_one \m@ne
\tex_let:D \c_minus_one \c_deprecateacro_minus_one
\ExplSyntaxOff
\makeatother
\usepackage{acro}

\renewcommand{\sp}{\mkern2mu}

\DeclarePairedDelimiter\ceil{\lceil}{\rceil}

\newcommand{\pkL}{\ensuremath{L}}
\newcommand{\gainR}{\ensuremath{\mathcal G_r}}
\newcommand{\gainC}{\ensuremath{\mathcal G_c}}
\newcommand{\gainRM}{\ensuremath{\bar{\gainR}}}
\newcommand{\gainCM}{\ensuremath{\bar{\gainC}}}

\newcommand{\Pfa}{\ensuremath{\mathsf P_{\mathsf{fa}}}}
\newcommand{\dm}{\ensuremath{\mathsf{d_m}}}
\newcommand{\interf}{\ensuremath{\mathcal I}}
\newcommand{\si}{\ensuremath{\mathbf x_1}}

\DeclareAcronym{CDF}{short = CDF ,long = cumulative distribution funtion}
\DeclareAcronym{HPBW}{short = HPBW ,long = half-power beamwidth}
\DeclareAcronym{mm-wave}{short = mm-wave ,long = millimeter wave}
\DeclareAcronym{OFDM}{short = OFDM ,long = orthogonal frequency-division multiplexing}
\DeclareAcronym{PPP}{short = PPP ,long = Poisson point process}
\DeclareAcronym{PRI}{short = PRI ,long = pulse repetition interval}
\DeclareAcronym{RCS}{short = RCS ,long = radar cross section}
\DeclareAcronym{SIR}{short = SIR ,long = signal-to-interference ratio}

\begin{document}

\title{Performance of Radar and Communication Networks  Coexisting in Shared Spectrum Bands}

\author{Andrea Munari, Nina Grosheva, Ljiljana Simi\'{c}, Petri M\"{a}h\"{o}nen
\thanks{
Authors are with the Institute for Networked Systems of the RWTH Aachen University, D-52072 Aachen, Germany (contact e-mail: nina.grosheva@rwth-aachen.de).}
}

\maketitle

\begin{abstract}
Recent technological advancements are making the use of compact, low-cost, low-power mm-wave radars viable for providing environmental awareness in a number of applications, ranging from automotive to indoor mapping and radio resource optimisation. These emerging use-cases pave the road towards  networks in which a large number of radar and broadband communications devices coexist, sharing a common spectrum band in a possibly uncoordinated fashion. Although a clear understanding of how mutual interference influences radar and communications performance is key to proper system design, the core tradeoffs that arise in such scenarios are still largely unexplored.
In this paper, we provide results that help bridge this gap, obtained by means of an analytical model and extensive simulations. To capture the fundamental interactions between the two systems, we study mm-wave networks where pulsed radars coexist with communications devices that access the channel following an ALOHA policy. We investigate the effect of key parameters on the performance of the coexisting systems, including the network density, fraction of radar and communication nodes in the network, antenna directivity, and packet length.  We quantify the effect of mutual interference in the coexistence scenario on radar detection and communication network throughput, highlighting some non-trivial interplays and deriving useful design tradeoffs.
\end{abstract}


\IEEEpeerreviewmaketitle

\section{Introduction} \label{sec:intro}

The past few years have witnessed a steadily increasing research attention towards the use of radars as enablers for environmental awareness in a number of emerging scenarios. The precursor of this trend has been the automotive domain, where radars are already employed in commercial solutions for obstacle detection and cruise control \cite{Hasch12_MTT,Kandeepan17_ITS,Heath18_TVT}. In turn, radar sensing capabilities in the mm-wave unlicensed frequency band are now also being considered for indoor mapping and enhanced localisation in 5G and Internet-of-Things use-cases \cite{Guidi16_TMC}, as well as to provide environmental awareness to support radio resource optimization, e.g. for beam-steering decisions in directional communications systems \cite{Simic16_RadCom}.

Such new applications are rendered possible by recent technological advancements, which have made compact and low-cost radar transceivers a reality, and trigger a fundamental change in paradigm. Namely, in contrast to traditional settings where radars transmit over dedicated and strictly regulated spectrum (e.g. in meteorology or aeronautics), the pervasive use of radars to support a multitude of new use-cases calls for operation over unlicensed bands shared with other devices and services. A relevant example in this direction is offered in \cite{Heath18_TVT}, where an IEEE 802.11ad-based system is proposed to support jointly broadband data links and radar detection in vehicular networks operating in the $60 \, \rm{GHz}$ band. Similar approaches are being explored also for drones and unmanned equipment \cite{Garmatyuk08_MilCom,Zhu15_Mobicom}, while the concept of a mm-wave personal mobile radar embedded in everyday communications devices such as mobile phones, laptops or tablets was proposed in \cite{Guidi16_TMC}.
It is then safe to assume that, in the near future,  heterogeneous networks where a large number of communications and radar transceivers share a common bandwidth in a possibly uncoordinated way to offer different and complementary services will be the norm \cite{Mehrnoush17_TCCN,Khawar14}.

These emerging scenarios pose new interesting challenges in terms of network and protocol design.  For communication nodes, radar pulses represent an additional source of interference, the characteristics of which may differ significantly from the disturbance generated by concurrent data links the system is devised to cope with. Namely, a radar may follow a different -- if any -- medium access policy compared to communications terminals, and its transmissions may take the form of periodic short bursts, in contrast to the typically longer and less regular channel occupation resulting from data packet exchanges. For the radars in such a spectrum-sharing network, performance is no longer influenced solely by noise or device capabilities, but can be severely affected by interference coming from other transmitters that may trigger undesired false alarms or impede proper target detection and tracking.

Understanding the impact of mutual interference on the two systems in complex topologies is thus paramount for proper design of upcoming radar-communications services, as well as for gaining insights on how they can scale in practical setups. However, a thorough characterisation of the design and performance tradeoffs in radar-communications networks is missing from the literature, which thus far has mainly focused on simple scenarios where few terminals share the same bandwidth  \cite{Zheng17_TSP,Mehrnoush17_TCCN,Nartasilpa16}.

As a step towards bridging this gap, we study in this paper the performance of large planar radar-communications networks. To derive insights that are not bound to specific protocol implementations but rather capture the basic and fundamental effects of spectrum sharing between the two types of transmissions, we focus on a scenario in which radars operate in pulse mode while communications devices  follow an ALOHA medium access policy. With an eye on applications in the mm-wave domain, we consider line-of-sight channels and the use of directional antennas for both radar detection and data exchange, possibly characterised by different antenna patterns. For the sake of  clarity and brevity, and in view of the well-understood fundamental behaviour of random access wireless systems,  we characterise the coexistence performance of communications devices by means of extensive network simulations. Importantly, we characterise analytically the performance of radars in the coexistence scenario using stochastic geometry tools. Specifically, we develop a simple and tractable analytical model based on the strongest interferer approximation that effectively captures the detectable range of radars in our heterogeneous setup. The derived closed-form expressions, whose accuracy is supported by simulation results, pinpoint the role of some key system parameters, offering useful design hints, and highlight the important role played by the spatial distribution of nodes.

We investigate the effect of different packet durations for communication links, and vary the density of devices as well as the fraction of radar and communications nodes in the network, so as to thoroughly capture the effect of interference on both systems. Our results show that, although spectrum sharing among uncoordinated radars and communication devices in large networks is viable, the role of mutual interference must be carefully evaluated and considered for proper system design. Our work thus represents a first step in identifying the key, yet still unexplored, tradeoffs of these heterogeneous networks, with the aim of stimulating further research in this area, considering e.g. more advanced medium access policies for radar such as the IEEE 802.11-like CSMA.

The rest of this paper is organised as follows. Section~\ref{sec:relWork} reviews the related work. Section~\ref{sec:sysModel} describes our system model. Section~\ref{sec:analysis} presents our analytical characterisation of radar performance. Section~\ref{sec:methodology} details our simulation methodology. Section~\ref{sec:results} presents our results of radar and communication network performance in a range of coexistence scenarios. Section~\ref{sec:concl} concludes the paper.

\section{Related Work}\label{sec:relWork}

Whereas the performance of communication networks in the presence of interference is thoroughly understood, little attention has been devoted to the behaviour of radars when competing transmissions are in place. Radar-to-radar interference studies have traditionally focused on simple  topologies with two devices sharing the same band  \cite{Foreman91}, \cite{Foreman95_TEC}, \cite{Brooker07_TEC} and have only recently been extended to larger networks. In \cite{Jondral13_EURASIP}, analytical bounds are presented for \ac{OFDM} radars modelling the aggregate interference as a Gaussian random variable. While insightful, this approximation cannot capture the role of the spatial distribution of nodes in the network. A relevant step forward was taken in \cite{Kandeepan17_ITS,Munari18_CL}, where the authors employ stochastic geometry tools to analyse the detectable range in large linear and planar topologies.

Regarding coexistence between radar and communications systems, prior works mainly target traditional use-cases in which radars transmit at higher power and enjoy primary access to the spectrum. Following this approach, information theoretic bounds on the achievable performance were derived in \cite{Bliss16_TSP} for a two-link topology. In a similar setup, \cite{Zheng17_TSP} provides optimised radar and communications design guidelines considering the structure of mutual interference. The effect of radar transmissions on communications links  is also studied in \cite{Nartasilpa16} and \cite{Mehrnoush17_TCCN}. In the former work, the performance degradation induced by a Gaussian-approximated radar interference on an uncoded communication system is determined, while \cite{Mehrnoush17_TCCN} proposes modifications to a Wi-Fi receiver that enable faster detection of radar signals and reduce their impact on broadband data links. Interference mitigation techniques based on a cognitive spectrum sharing approach between LTE and radars are explored in \cite{Khawar14}. Finally, tradeoffs emerging in networks where individual terminals switch between radar detection and communications operations are tackled in  \cite{Ren18_WCL}.

To the best of our knowledge, this paper is the first extensive study of coexistence and mutual interference between low-power radars and communication devices in large and uncoordinated planar networks, which is key for understanding and engineering emerging scenarios where pervasive radar nodes share e.g. mm-wave unlicensed spectrum with broadband communication nodes.

\section{System Model} \label{sec:sysModel}

\subsection{Network Model}
\label{sec:network}
We consider a planar network in which communication devices (a fraction $ \beta $ of the overall nodes) share the same spectrum band with pulsed radars (fraction $ 1- \beta $). The positions of the devices are modelled according to a \ac{PPP} $\Lambda \subset \mathbb R^2$ of intensity $\lambda$. For the sake of tractability, we assume time to be divided in slots of equal duration $ T_s = 1/B $, where $ B $ is the available bandwidth, and all nodes to be slot-synchronous.

Each radar tries to detect a target with a \ac{PRI} of $M$ timeslots. Namely, it transmits a pulse in one timeslot and then switches to receiver mode for the next $M-1$ timeslots, waiting for an echo. To capture the uncoordinated nature of the system under study, random offsets among the operating cycles of different radars are assumed. Specifically, a radar device $\mathbf x_i \in \Lambda$ transmits its pulses at slots $ \nu_i + nM, n\in \mathbb{N}$, where $\nu_i$ is a mark for the PPP independently drawn for each node from a uniform distribution, i.e. $ \nu_i \sim \mathcal U \left\{0,M-1 \right\} $. It follows that, on average, a fraction $ 1/M $ of the radars are active in each slot.

\begin{figure}\begin{center}
\includegraphics[width=.75\columnwidth]{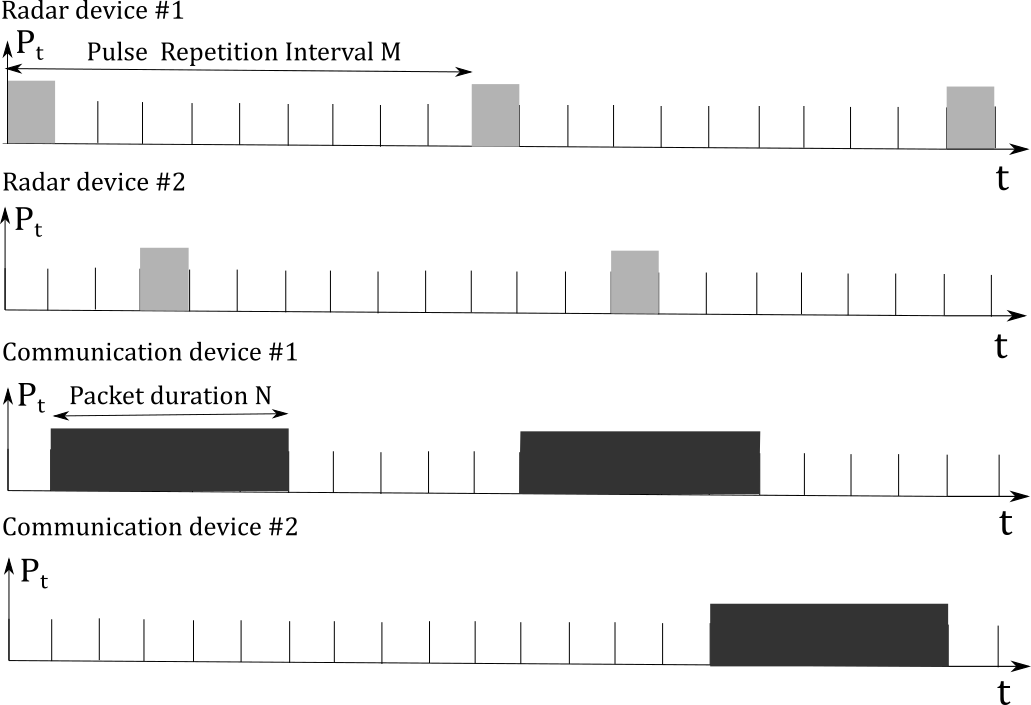}
\caption{Example timeline for the operation of two radars and two communication devices in the considered system model.}
\label{fig:timeline}
\end{center}\end{figure}

A communication device, instead, follows a $p_t$-persistent ALOHA policy to send packets of duration $\pkL$ timeslots to a dedicated receiver, randomly located over a circle of radius $ d_c $  centred at the sender. As for the radars, a random offset among the operational cycles is allowed, so that a communication node $\mathbf x_j \in \Lambda$ decides at slots $\nu_j + nM$ (with $n\in \mathbb N$ and $ \nu_j \sim \mathcal U\left\{0,M-1 \right\}$) whether to transmit a packet (with probability $ p_t $), or to defer access for the subsequent $\pkL$ slots prior to attempting again (with probability $ 1-p_t $).  An example timeline for the mixed radar and communication network is shown in Fig.~\ref{fig:timeline}.

\subsection{Channel and Interference Model}
\label{sec:channel}

All devices use the same operating frequency $ f_{c} $ and bandwidth $B$, and employ the same transmit power $ P_{t} $. Communications and radar nodes are equipped with an antenna of radiation pattern $\gainC(\psi) $  and $\gainR(\psi), \psi\in \left[ 0,2\pi \right) $, respectively.  Focusing on mm-wave operation, we assume line-of-sight propagation, considering a power-law path loss of exponent $\alpha=2$ and no fading. A receiving device is then affected over a slot of interest $ \ell $ by an aggregate interference
\begin{equation}
\interf_{\ell} = \sum_{i} P_t  \,\mathcal G_{tx} \mathcal G_{rx} \kappa \,{d_i}^{-\alpha} .
\label{eq:interference}
\end{equation}
In \eqref{eq:interference}, $\kappa := \left[c /(4 \pi f_c )\right]^2$, the summation is taken over all nodes that are transmitting in timeslot $ \ell $, $\mathcal G_{tx}$ and $\mathcal G_{rx}$ are the transmit and receive gain dictated by the type of interfering and receiving devices as well as by their relative orientation, $ c $ is the speed of light, and $ d_i $ is the interferer-receiver distance.

For a communications link, we assume that transmitter and receiver are pointing towards each other, so that the useful signal at the communication latter takes the form
\begin{equation}
S_c = P_t \gainCM^2  \, \kappa \, {d_{c}}^{-\alpha},
\label{eq:signal-comm}
\end{equation}
where $d_{c}$ is the distance between the transmitter and receiver and $\gainCM$ is the maximum gain of the antenna pattern of the communication devices.

For a radar node, the useful signal is instead the incoming reflection  of the transmitted pulse generated by a target. Assuming the target to be located at a distance $ d_r $ from the radar and along its boresight direction, the received power of the echo follows from the radar equation as
\begin{equation}
S_r = \frac{P_t \gainRM^2 \kappa\, \sigma \,G_p }{4\pi} \cdot \,d^{-2 \alpha}_{r},
\label{eq:signal}
\end{equation}
where $ \sigma $ is the \ac{RCS} and is a measure of the reflectivity of the target,\footnote{In order to better isolate the effect of interference, we consider a fixed RCS, representative of static targets.} $ G_p $ is the sensitivity gain from further signal processing and $\gainRM$ is the maximum gain of the antenna pattern of the radar devices.
Due to the interference-limited nature of the networks under study, and so as to better stress the mutual interplay of radars and communications transmissions, we disregard the effect of noise.\footnote{Following stochastic geometry arguments, the impact of noise on communications links would appear as a scaling factor on throughput at very low densities, not altering the trends we report. As for radars, additional results, not shown here due to space constraints, confirm that thermal noise also becomes non-negligible only in very sparse networks, while for all densities considered in our study mutual interference is the key performance driver (see also \cite{Munari18_CL}).}

\subsection{Performance Metrics}
\label{sec:performance}
\textbf{Radar devices:} Radar devices implement a detection rule based on incoming power, declaring a target to be present if the overall power over a slot exceeds a threshold $ \theta $. A target is then correctly detected over a slot $ \ell $ if $ S_r + \interf_{\ell} > \theta $. Conversely, if -- in the absence of a target -- the interference level in one or more of the $ M-1 $ timeslots during which the radar is listening for the echo is strong enough to cross the detection threshold, a false alarm is triggered. Following common radar design principles, the detection threshold $ \theta $ is set according to the level of interference in the network, so that the probability of false alarm is kept at an acceptable level.

We evaluate the performance of the radars based on their detectable range by determining the maximum distance $\dm $ at which targets can be reliably (i.e. with probability $1$) detected. Due to the deterministic nature of the target echo
given in~\eqref{eq:signal}, $\dm$ can be computed by setting $ S_r(\dm) = \theta $, to obtain
 \begin{equation}
\dm = \left(\frac{P_t \gainRM^2 \, \kappa  \sigma \, G_p } {4 \pi \theta}\right) ^{1/2\alpha}.
 \label{eq:dm}
 \end{equation}

\textbf{Communication devices:} A communication packet is considered to be successfully decoded if the average \ac{SIR} $\gamma$ at the receiver exceeds a threshold $ \Gamma $. Assuming capacity achieving codes, the decoding condition can thus be expressed as $\gamma > 2^R - 1$, where $ R $ is the information bitrate in bit/symbol. In turn, the average \ac{SIR} for a packet starting at slot $ \ell_0 $ can be computed as
\begin{equation}
\gamma = \frac{1}{\pkL}\sum_{\ell=\ell_0}^{\ell_0+\pkL} \dfrac{S_c}{\interf_\ell},
\label{Sir-m}
\end{equation}
where $ S_c $ and $ \interf_{\ell} $ are given by~\eqref{eq:signal-comm} and~\eqref{eq:interference}, respectively.
We evaluate the performance of the communication nodes based on the aggregate throughput density $ \tau $, expressing the average number of bits per second successfully exchanged per unit of area in the network. Assuming a symbol time $ 1/B $, we get
 \begin{equation}
 \tau = \lambda \beta p_t  \, R B \, \mathsf{p_s} , \label{throughput}
 \end{equation}
where $ \mathsf{p_s} = \mathbb P\left\{ \gamma > \Gamma \right\}$ is the probability of successfully delivering a packet.

\section{Analysis} \label{sec:analysis}

The performance of radar and communications devices in the heterogeneous networks under study is fundamentally driven by the aggregate level of interference. However, despite the conceptually simple definition in \eqref{eq:interference}, no closed-form solution for \interf\ is known in the considered setting with dominant line-of-sight components (i.e. no fading, and $\alpha \simeq  2$), and existing integral expressions are often involved and possibly difficult to be evaluated numerically \cite{DiRenzo15_TWC,Sousa90_JSAC,Zorzi18_arXiv}. In view of this, we characterise the coexistence performance of communications devices by means of detailed network simulations in Sec.~\ref{sec:results}. For radars, we follow a different approach and derive a tractable analytical model under the assumption that the performance of a node is solely dictated by its strongest interferer.\footnote{From well-known stochastic geometry results, applying the strongest interferer approximation to communications links would only offer a loose and scarcely insightful upper bound to the achievable performance, see e.g. \cite{Haenggi_book}.}  Such an approximation, whose tightness will again be verified via simulations, leads to simple expressions that reveal some fundamental and non-trivial tradeoffs on the coexistence of radars and communications devices.

Let us then focus without loss of generality on the typical radar node $\mathbf x_0$, located at the origin of the plane and emitting pulses at slots $nM$, $n\in \mathbb N$ (i.e. with mark $\nu_0 = 0$). Furthermore, indicate as $\si \in \mathbb R^2$ the coordinates of its strongest interferer. Disregarding the activity of other devices, a false alarm occurs with probability 
\begin{equation}
\Pfa = \mathbb \pi_a \cdot  \,\mathbb P\{ P_t  \mathcal G_{tx} \mathcal G_{rx} \kappa \left\| \si \right\|^{-\alpha}  \geq \theta\}
\label{eq:pfa}
\end{equation}
where $\pi_a$ denotes the probability that the strongest interferer transmits at least once during the $M-1$ slots spent by the typical radar waiting for a target echo, whereas the second factor accounts for whether the generated interference is high enough to trigger an erroneous detection. In turn, $\pi_a$ depends on the channel access pattern followed by the interferer. If \si\ is a radar node, it will transmit exactly once over the $M-1$ slots of interest as soon as $\nu_1 \neq 0$, i.e. with probability $1-1/M$. If, \si\ is a instead communications device, the probability of triggering interference takes the form $1-(1-p_t)^{\omega(\nu_1)}$, where $\omega(\nu_1)$ indicates the number of packets, i.e. of transmission opportunities for the node, that overlap with the echo waiting time for the typical radar. Recalling that a fraction $\beta$ of the devices operate in communications mode and that $\nu_1 \sim \mathcal U\{0,M-1\}$, we can then write
\begin{equation}
\pi_a = (1-\beta) \left( 1- \frac{1}{M} \right) + \frac{\beta}{M} \sum_{\nu_1=0}^{M-1} \left(1-(1-p_t)^{\omega(\nu_1)}\right)
\label{eq:pi_a}
\end{equation}
where $\omega(\nu_1)$ follows by simple counting arguments reported in the Appendix as
\begin{align}
\begin{split}
\omega(\nu_1) = 1 &+ \max\left\{ 0, \ceil*{\frac{\nu_1 - 1}{L}}  \right\} \\
&+ \ceil*{ \frac{M-1 - \min\left\{ \nu_1+L-1, M-1 \right\}}{L} } .
\end{split}
\label{eq:numPkts}
\end{align}

Let us consider now the second factor in \eqref{eq:pfa}, expressing the probability for a transmission from \si\ to induce a false alarm, and assume for the sake of tractability an ideal antenna pattern for all devices, with gain $\bar{\mathcal G}$ within a beam of width $\varphi$ and $0$ elsewhere. In this case, only nodes whose antenna pattern overlaps with that of the typical radar are received with a non-zero gain, and, by virtue of the independent orientation of boresight directions, their positions are captured by a thinned version $\Lambda'$ of the original PPP, of intensity $\lambda'=\lambda [(\varphi/(2\pi)]^2$. The strongest interferer is then simply the device belonging to $\Lambda'$ closest to the origin of the plane, so that
\begin{equation}
\mathbb P\left\{ \left\| \si \right\| > y \right\} = \mathbb P \left\{ \Lambda' \cap b(0,y) = \emptyset \right\} = e^{-\lambda' \pi y^2}
\label{eq:cdfStrongest}
\end{equation}
where $b(0,y)$ denotes the circular region of radius $y$ centred at the origin.\footnote{Strictly speaking, \eqref{eq:cdfStrongest} holds for $\alpha>2$. Reported results shall then be interpreted as obtained for  $\alpha \rightarrow 2$.} Substituting this result into \eqref{eq:pfa} readily leads to 
\begin{equation}
\Pfa = \pi_a \cdot \left( 1 - \exp\left( -\frac{\lambda \varphi^2}{4\pi} \left( \frac{P_t \sp \bar{\mathcal G}^2 \sp \kappa}{\theta} \right)^{2/\alpha} \right) \right ).
\label{eq:pfa_exp}
\end{equation}
Finally, solving \eqref{eq:pfa_exp} with respect to $\theta$ allows us to compute the detection threshold and, substituting the obtained result in \eqref{eq:dm}, the maximum detectable range under the strongest interferer approximation follows as:
\begin{equation}
\mathsf{d_{m,a}} = \left( \frac{\sigma G_p}{4\pi} \right)^{1/2\alpha} \left( -\frac{\lambda \varphi^2}{4\pi \ln \left( 1- \Pfa/\pi_a \right)} \right)^{1/4}.
\label{eq:dma}
\end{equation}

The closed-form formulation derived for $\mathsf{d_{m,a}}$ offers a simple yet powerful tool to predict radar performance in the complex heterogeneous setups under study. Namely, the expression in \eqref{eq:dma} conveniently isolates the impact of distinct transmission patterns and fraction of radar and communication devices -- embedded in $\pi_a$-- , from the effect of topological, radar, and antenna parameters. Moreover, as will be discussed in depth in Sec.~\ref{sec:results}, \eqref{eq:dma} highlights and quantifies the key role played by the network density, while revealing that, in an interference-limited setting, the detectable range does not depend on the employed transmission power, operating frequency or bandwidth.

\section{Simulation Methodology} \label{sec:methodology}

To verify the accuracy of the analytical model developed in Sec.~\ref{sec:analysis} and to study the coexistence performance of communication nodes, we performed extensive Monte Carlo simulations. In order to examine a wide range of topologies for the radar-communications network, we consider a total node density $ \lambda $, from $ 10^{-5} $ to \unit[$10^0 $] {devices / m$^2$}, with communication nodes accounting for $ \beta=0\% $, $33\%, 66\%$ or $100\% $ of the total nodes in the heterogeneous network. We varied the simulated network area according to the density, so as to ensure that the number of nodes is sufficient to obtain statistically stable outcomes. All reported results are averaged over 100 \ac{PPP} realisations, each studied for a duration of 6000 timeslots.

We focus on operation in the mm-wave band, setting $f_c = 60$ GHz.\footnote{This licence-free band is particularly interesting for emerging radar/communications coexistence, as low-cost CMOS chipsets operating at these frequencies are starting to be available \cite{Guidi16_TMC}.} All devices employ directional antennas with a reference \ac{HPBW} of approximately $30^\circ$. The antenna pattern was generated using the \mbox{\textsc{MATLAB}} \emph{Phased Array System Toolbox} to model a uniform quadratic antenna array with a perfectly absorbing backplane. Specifically, a $4\times 4$ array of isotropic antennas spaced by half wavelength was considered. The resulting pattern has a maximum gain of $16.5 \,\rm{dBi}$ for the main lobe, with an attenuation of $11.3 \,\rm{dBi}$ for the maximum gain of the sidelobes.  Additionally, we also consider a case where the radar devices use less directional antennas with a \ac{HPBW} of approximately $60^\circ$. The size of the array in this case is $2\times2$ and the maximum gain of the main lobe is 10.1~dBi.

We choose a target false alarm probability $\Pfa = 0.1$ for radar nodes.\footnote{This value of  $\Pfa$ can be representative of non safety-critical radar use-cases that characterise emerging applications, e.g. environmental awareness, where a higher rate of false detections may be permissible. Nonetheless, we note that our methodology and the qualitative trends of our simulation results hold also for more stringent values of \Pfa.} In order to determine the detection threshold $\theta$, the total power seen by radar nodes in the absence of a target -- determined by the positions and transmission pattern followed by other transmitters -- is computed for every PPP realisation over each timeslot. For a given value of  $\theta$, the expected false alarm rate is then computed by counting the average number of pulse repetition intervals during which the receiver experiences a power larger than $\theta$ in at least one of the slots spent listening for a target echo. By iteratively testing different values of $\theta$, our simulations determine the threshold that provides the desired average false alarm probability.

For communication nodes, the reference duration for a data packet is half of a radar \ac{PRI} $ M $, i.e. $\pkL=30$ timeslots. All other relevant simulation parameters are reported in Table~\ref{tab:parameters}.

\begin{table}
	\renewcommand{\arraystretch}{1.25}
	\centering
    \caption{Simulation parameters}
	\begin{tabular}{ l | l l}
		\hline
		\multicolumn{2}{l}{\textsc{Parameter}} & \textsc{Values}\\
		\hline
		\hline
		$\lambda$& density of devices & \unit[$10^{-5} - 10^0 $] {device/ m$^2$} \\
		$\beta$& percentage of comm devices & 0\%, 33\%, 66\%, 100\%\\
		$P_t$& transmission power & \unit[10]{dBm} \\
		$f_c$& operational frequency & \unit[60]{GHz}\\
		$\alpha$& path-loss exponent & 2\\
		$\varphi$& reference \ac{HPBW} & $30^\circ$\\
		$B$& signal bandwidth & \unit[300] {MHz}\\
		$M$& radar \ac{PRI}  & 60 timeslots \\
		$G_p$& radar processing gain & 10 \\
		$\sigma$& \ac{RCS} of the target & 10 m$^2$\\
		$\Pfa$& targeted probability of false alarm & 0.1 \\
		$p_t$& tx persistency, comm devices & 0.1 \\
		$R$& transmission rate & \unit[$1.5$] {bits/symbol} \\
		$d_{c}$& tx -- rx distance & 55 m\\
		$\pkL$& reference packet duration & $ M/2 = 30 $ timeslots\\
		\hline
	\end{tabular}\label{tab:parameters}
\end{table}

\section{Results} \label{sec:results}

\subsection{Effect of Network Density}
\label{sec:density}
 We begin our analysis focusing on the reference case in which data packets have a duration of $ \pkL = M/2 = 30 $ slots and communication and radar devices employ the same antenna pattern with \ac{HPBW} of $ \varphi = 30^\circ $. Such a symmetric setup allows us to better isolate the role played by the different transmission patterns of radar and communication operations, deriving first insights on the effect of mutual interference.

\begin{figure}
\centering
\includegraphics[width=.8\columnwidth]{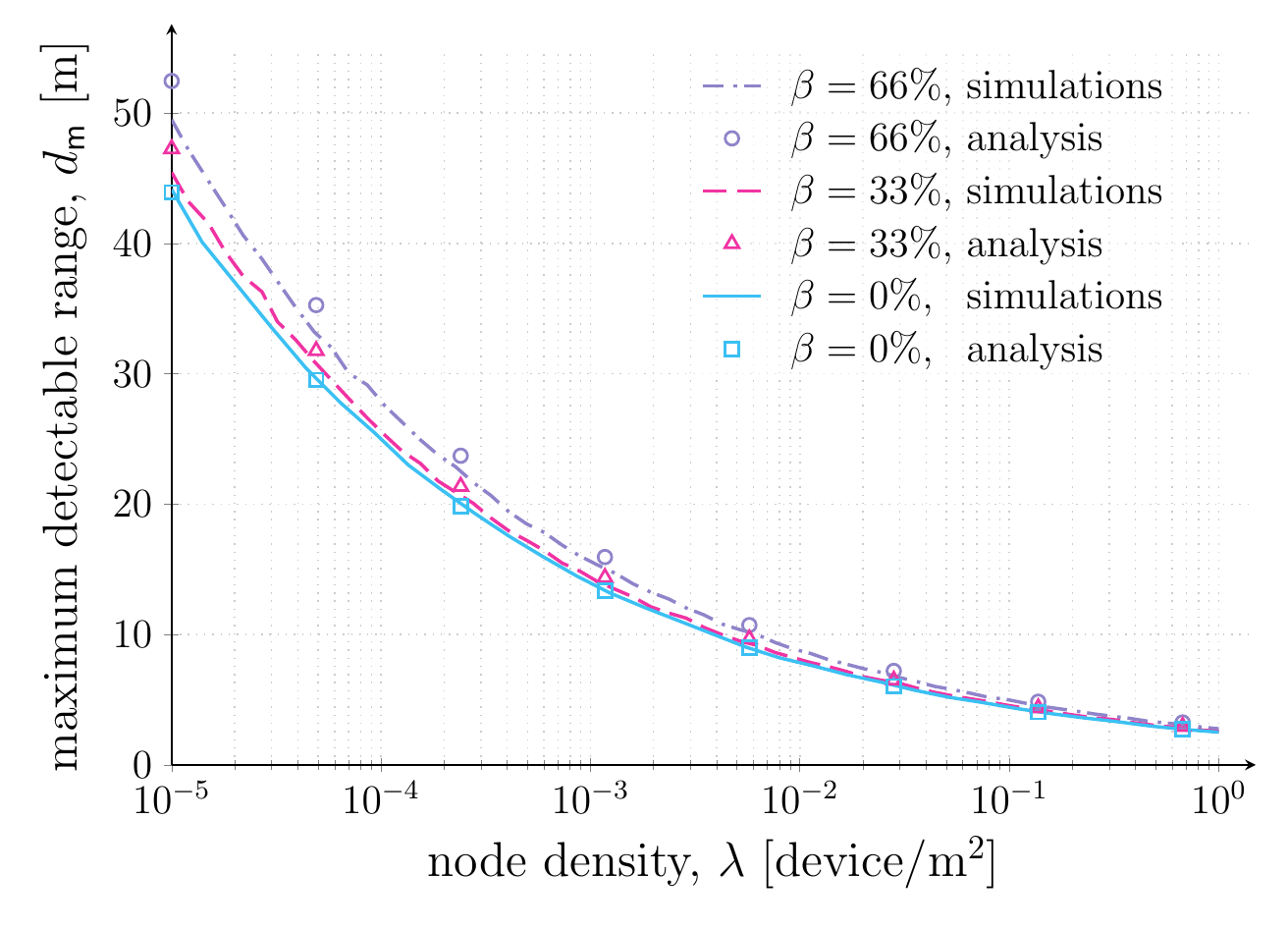}
\vspace{-4mm}
\caption{Maximum detectable radar range vs. total network density, with different fractions $\beta$ of communication devices. Markers show analytical results under the strongest interferer approximation and lines show simulation results; $ \varphi = 30^\circ $, $ \pkL=M/2 $ slots.}
\label{fig:dm_30_30}
\end{figure}

Fig.~\ref{fig:dm_30_30} presents radar performance in terms of the maximum detectable distance $\dm$ against the total network density $ \lambda $, for different fractions $\beta$ of communication nodes. In the plot, lines report simulation results, i.e. considering a realistic antenna pattern and aggregate interference, while markers indicate the detection range $\mathsf{d_{m,a}}$ predicted by the analytical model based on the strongest interferer approximation and ideal antenna pattern developed in Sec.~\ref{sec:analysis}.
The significant impact of interference is apparent in the exponential drop of the maximum detectable distance experienced in denser networks. In sparse topologies, often encountered in traditional applications where radars are granted priority access to the spectrum, targets can be reliably detected over large distances. Conversely, as we increase the density of devices sharing the same channel, performance significantly deteriorates, until in very densely deployed networks only targets within a few meters of the radar can be detected. This result provides a first interesting design hint, offering a quantitative trend on the achievable radar range, and suggesting that the design of medium access policies for radar operations will be needed for emerging applications in shared spectrum.

\begin{figure}
\centering
\includegraphics[width=.8\columnwidth]{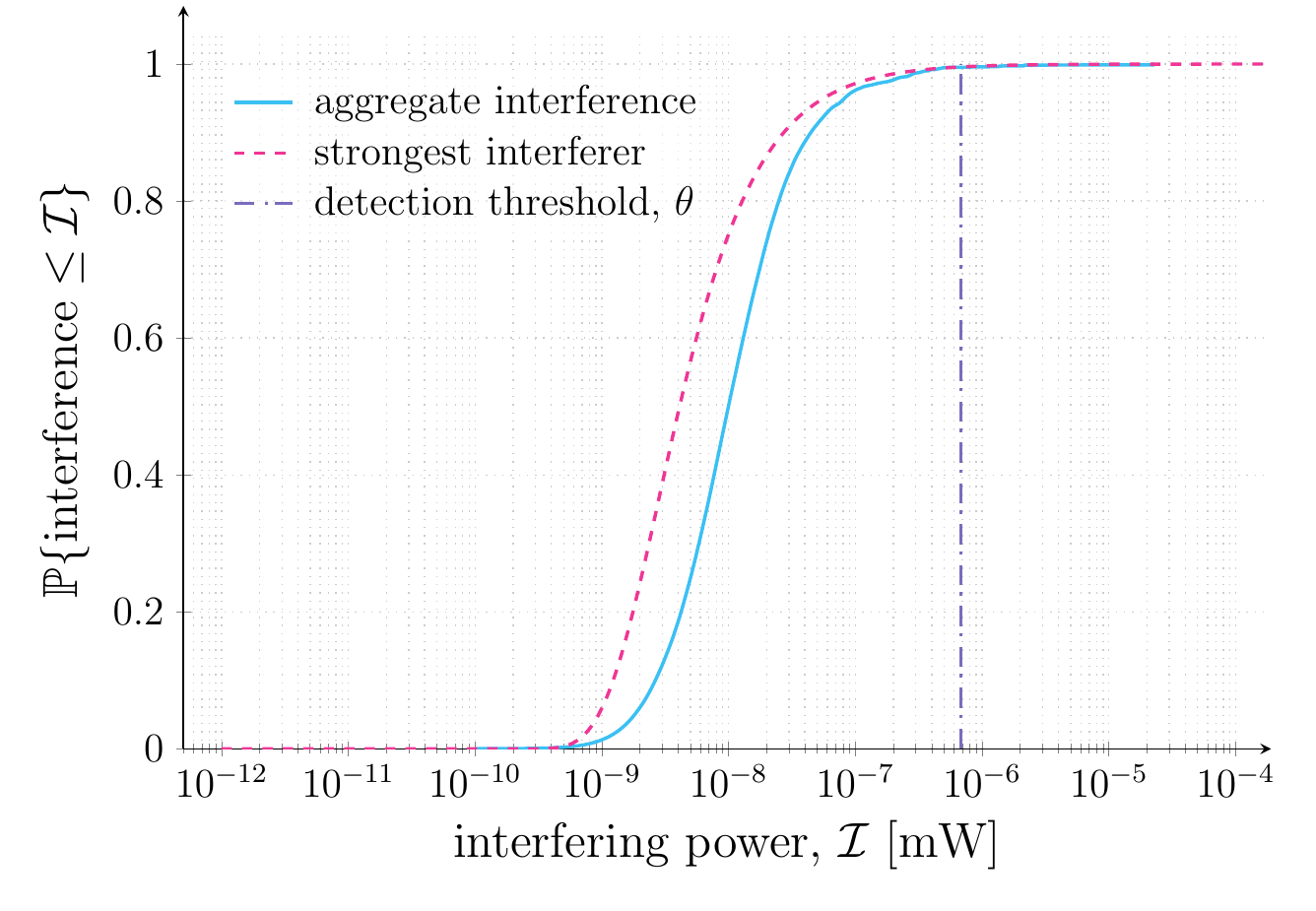}
\vspace{-4mm}
\caption{CDF of the aggregate interference and the strongest interferer (simulation results); \mbox{$ \lambda = 10^{-3} $} device/m$ ^{2} $, $ \beta = 33\% $, $ \varphi = 30^\circ $, $ \pkL=M/2 $ slots. }
\label{fig:cdf}
\end{figure}

Fig.~\ref{fig:dm_30_30} also highlights a very tight match between our analytical and simulation results, supporting the validity of the presented model, and confirming that the main cause of false alarms is not the aggregate interference from the whole network, but rather the presence of a \emph{single} device close to a radar which constantly disrupts its detection. This behaviour is confirmed by Fig.~\ref{fig:cdf}, which compares -- for a reference density $\lambda=10^{-3}$ -- the cumulative distribution function of the \emph{aggregate} interference perceived by a radar  over a slot (solid line) to that of the interference caused by the strongest interferer active over that slot only (dashed line).  The vertical dashed line indicates the value of the detection threshold $\theta$ computed considering the aggregate interference to achieve the target false alarm rate. We can observe that in the region of interest for setting the threshold, both functions are nearly identical, indicating that the main limitation on system performance indeed comes from the strongest interferer.

Furthermore, Fig.~\ref{fig:dm_30_30} illustrates the influence of the communication system, revealing that an increase in the fraction of communication devices $\beta$ has little effect on radar performance, and can in fact be slightly beneficial. This trend might at first appear counterintuitive, as communication links generate a higher level of interference than radars. Namely, recalling that a persistency $p_t= 0.1$, and a radar \ac{PRI} of $60$ timeslots are considered, a communication device transmits approximately 10\% of the time, while a radar sends pulses only 6\% of the time. To explain this behaviour, let us then consider how the transmission pattern of the strongest interferer affects the rate of false alarms. If the strongest interferer is a radar device, it will transmit on a fixed schedule once every \ac{PRI} and cause a false alarm during every detection. On the other hand, if this device is a communication device, it will decide whether to transmit or not twice in each \ac{PRI} with a probability of transmission of 0.1. Therefore, some of the detections will remain unobstructed by the strongest interferer as the device remains idle during the detection. If the communication device chooses to transmit, it will occupy the channel for much longer than a radar would. This, however, does not have an effect on radar performance since a false alarm during a single slot is already enough to prevent the detection. Therefore, false alarms in multiple slots during a single detection attempt do not harm radar performance any more than a single one would. We can conclude that, although  communication devices occupy the channel for a longer percentage of time, their transmission activity is less detrimental to the radars, pinpointing a possible benefit for the coexistence of the two systems.

\begin{figure}
\centering
\includegraphics[width=.8\columnwidth]{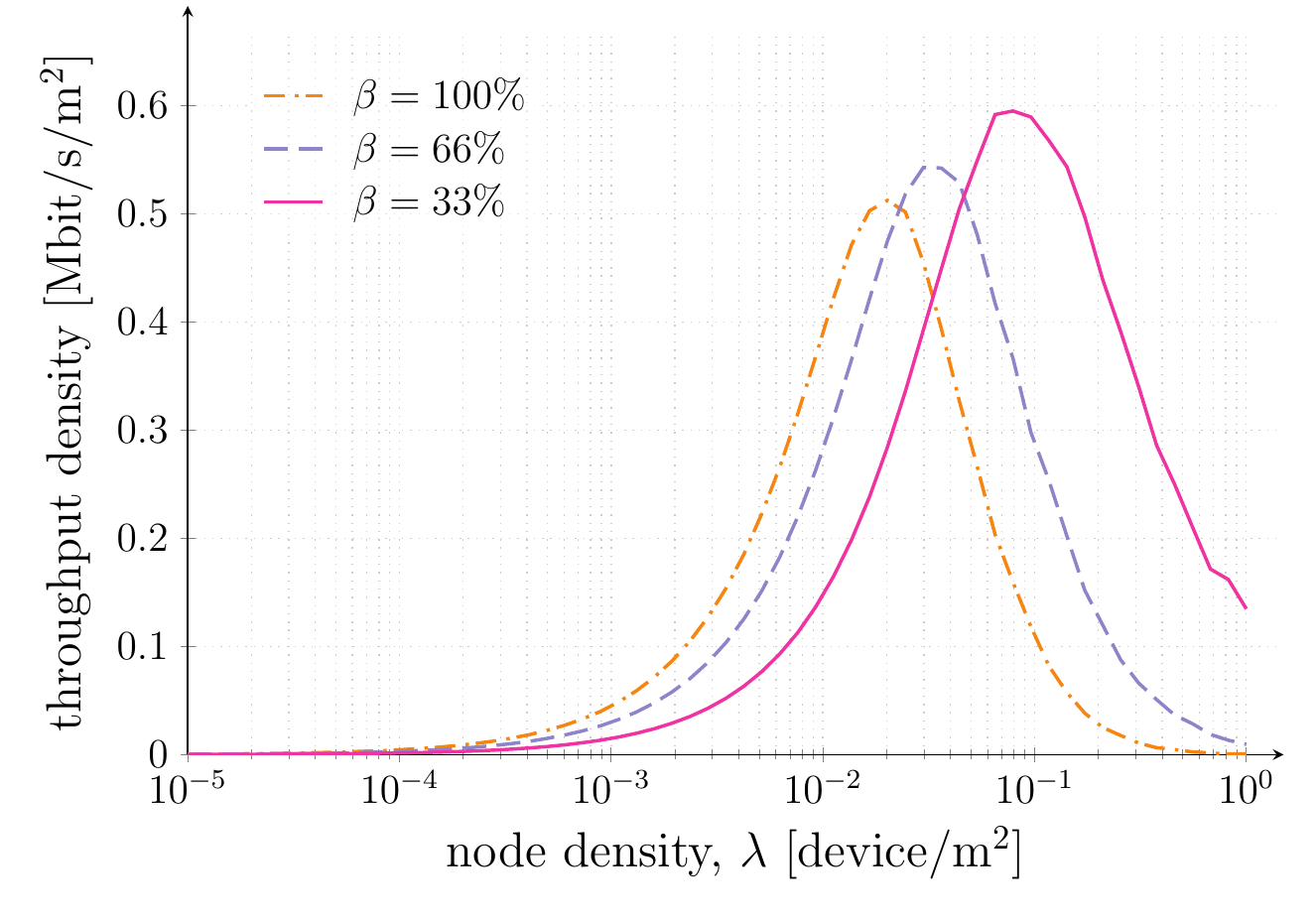}
\vspace{-4mm}
\caption{Throughput density vs. total network density, with different fractions $ \beta $ of communication devices (simulation results); $ \varphi = 30^\circ $, $ \pkL=M/2 $ slots.}
\label{fig:t_30_30}
\end{figure}

To complement our discussion, we show in Fig.~\ref{fig:t_30_30} the communication performance, reporting the aggregate throughput density against the total node density of the heterogeneous network. As expected, the curve exhibits an ALOHA-like trend, reaching a maximum value for intermediate densities prior to dropping sharply due to excessive interference. Observing the effect of coexistence with a radar system, we can see that decreasing the percentage of communication devices shifts the throughput curve to the right, enabling operation in denser networks. Moreover, if we consider the effect of $ \beta $ on the maximum achievable throughput, the opposite trend emerges compared to that for radar performance. Namely, whereas larger values of $\beta$ were beneficial for the detectable radar range, they decrease the maximum throughput of the communication nodes. Both these effects indicate that communication performance is affected by the aggregate interference in the network rather than the strongest interferer, and therefore the lower transmission activity of the radars is beneficial for the communication system.  Furthermore, recalling that successful decoding of communication packets requires that the average \ac{SIR} during the transmission of the packet is above a threshold, the transmission activity of the radars is more benign for the communication devices. Radar interferers send pulses with the duration of a single slot and will interfere with a small part of the packet, in many cases not preventing successful decoding at the receiver. The significantly longer transmissions from communication interferers, by contrast, are more likely to cause failed transmissions as they can affect larger portions of the packet. Therefore, it appears that coexistence with a radars can also positively affect the performance of the communication nodes, due to the infrequent radar transmissions.

However, comparing Fig.~\ref{fig:dm_30_30} and~\ref{fig:t_30_30} we can observe that the radar and communication nodes operate optimally at different network densities, possibly hindering the viability of some coexistence scenarios. Radar operation is severely limited by interference and therefore sparse topologies are most suitable for detection over large distances. Communication systems, on the other hand, cannot achieve high aggregate throughput in such networks and operate much more efficiently in denser deployments. When considering coexistence between the two systems, there thus always exists a trade-off between radar range and aggregate communication throughput. Given this, coexistence may be more favourable also from the communications viewpoint for applications targeting high per-node rather than aggregate throughput.

\subsection{Effect of the Communication Packet Duration}
\label{sec:packet}

In order to further study the tradeoffs that arise in the heterogeneous network, we consider the effect of communications packets of different duration. For a fair comparison, the same medium access persistency $p_t=0.1$ is considered, not altering the total amount of interference in the network.

\begin{figure}
\centering
\includegraphics[width=.8\columnwidth]{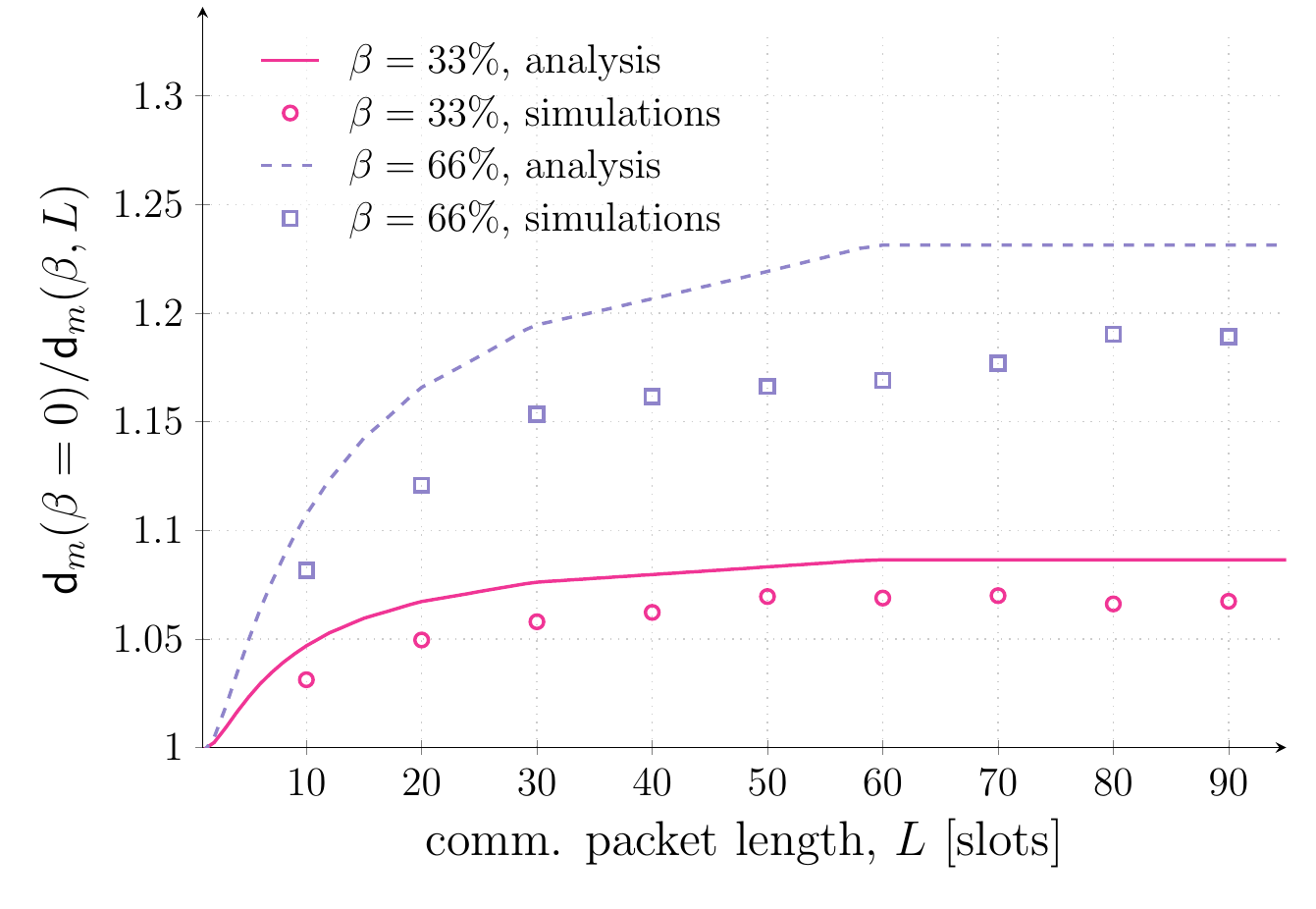}
\vspace{-4mm}
\caption{Ratio of radar detection range obtained under a given $(\beta,\pkL)$  configuration to the one achievable when a network of the same density is entirely composed of radar devices (i.e. for $\beta=0$) vs communications packet length \pkL.  Lines show analytical results, markers show simulation results for $\lambda=10^{-3}$. In all cases. $ \varphi = 30^\circ $.}
\label{fig:dmGain}
\end{figure}

To gauge the effect of communication links on radar performance, we compute the ratio $\Xi$ of the detectable range obtained under a given  configuration in terms of $(\beta,\pkL)$ to the one achievable when a network of the same density is entirely composed of radar devices (i.e. for $\beta=0$). Using the strongest interferer approximation and the formulation of $\mathsf{d_{m,a}}$ in \eqref{eq:dma}, this metric can be conveniently derived as
\begin{equation}
\Xi = \frac{\ln\big( 1 -  M \Pfa/(M-1) \big)}{\ln\big( 1 - \Pfa/\pi_a \big)}\,.
\label{eq:ratio}
\end{equation}
The expression in \eqref{eq:ratio} offers interesting insights, clarifying how the effect  on radars of sending communications packets of different duration is independent of the network density $\lambda$, and is rather defined by the channel access pattern of the devices, characterised by $\pi_a$, which, recalling \eqref{eq:pi_a} is a function of $\beta$ and $\pkL$. The trend exhibited by $\Xi$ for values of $\pkL$ ranging from $1$ to $95$ slots is reported in Fig.~\ref{fig:dmGain}. The plot confirms that the analytical curves are in good agreement -- within $4\%$ -- with the corresponding simulation results (circle markers).
Moreover, Fig.~\ref{fig:dmGain} clearly shows that increasing the duration of data packets exchanged among communications devices leads to an improvement in radar performance.\footnote{Note that, for any $L>M$, at most two transmission opportunities for a communication node overlap the echo waiting time for a radar node. In these conditions, $\pi_a$ evaluates to $(1-\beta)(1-1/M) + \beta [ 2 p_t + (M-2)(2p_t - p_t^2) ]/M$), regardless of $L$, leading to the saturation value visible in Fig.~\ref{fig:dmGain}.} Following the same line of reasoning discussed in Sec.~\ref{sec:density}, such a behaviour stems from the beneficial impact of having longer periods characterised by lower levels of interference when communications nodes defer channel access. Accordingly, the effect becomes more pronounced when a larger fraction of the network population operates in communications mode, as apparent by comparing the outcomes for $\beta=66\%$ (dashed line) and $\beta = 33\%$ (solid line). Notably, in the former configuration, an improvement in the order of $20\%$ in terms of detectable range is achievable, suggesting a convenient coexistence scenario from a radar standpoint.

For communications links, simulation results -- not reported due to space constraints -- revealed no significant change in throughput performance for different values of \pkL, regardless of  $\beta$. This trend, expected and well known in a purely ALOHA communications network, emerges also in the considered heterogeneous networks by virtue of two contrasting effects. On the one hand, recalling that echo detection pulses are emitted every $M$ slots, long data exchanges are prone to suffer interference from more radar transmissions. On the other hand, the longer a data packet, the smaller the detrimental impact of each interfering radar pulse on the average experienced SIR. The two factors balance each other out, resulting in a resiliency of communications links to the presence of radar interferers.

\subsection{Effect of Antenna Directionality}
\label{sec:directionality}

\begin{figure}
\centering
\includegraphics[width=.8\columnwidth]{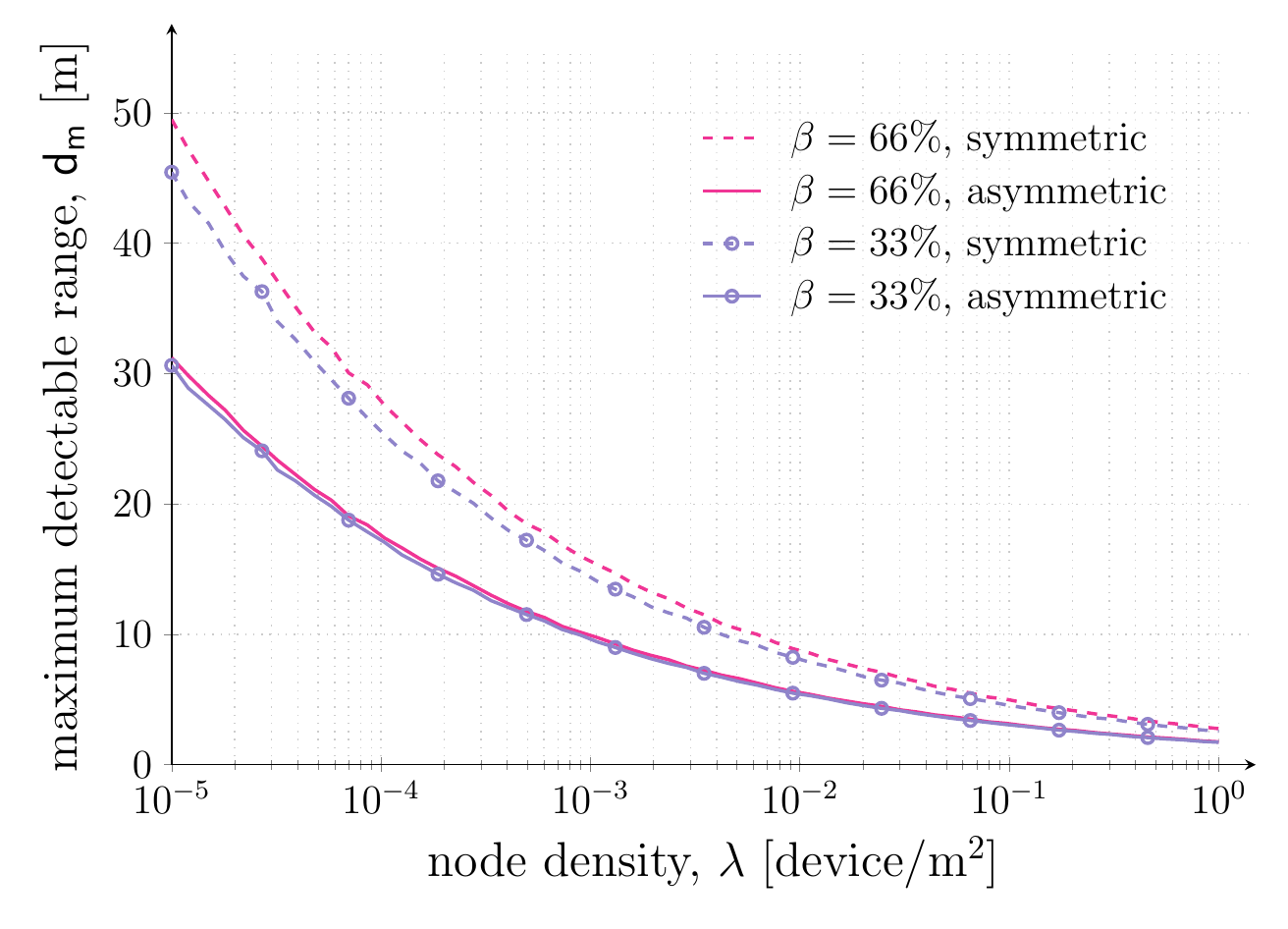}
\vspace{-4mm}
\caption{Maximum detectable radar range vs. total network density, with different fractions $ \beta $ of communication devices (simulation results). Packet duration $ \pkL=M/2 $ slots, $ \varphi_c = 30^\circ $, and $ \varphi_r = \{30^\circ, 60^\circ\}$ in symmetric and asymmetric antenna configurations, respectively.}
\label{fig:dm_60_30}
\end{figure}

\begin{figure}
\centering
\includegraphics[width=.8\columnwidth]{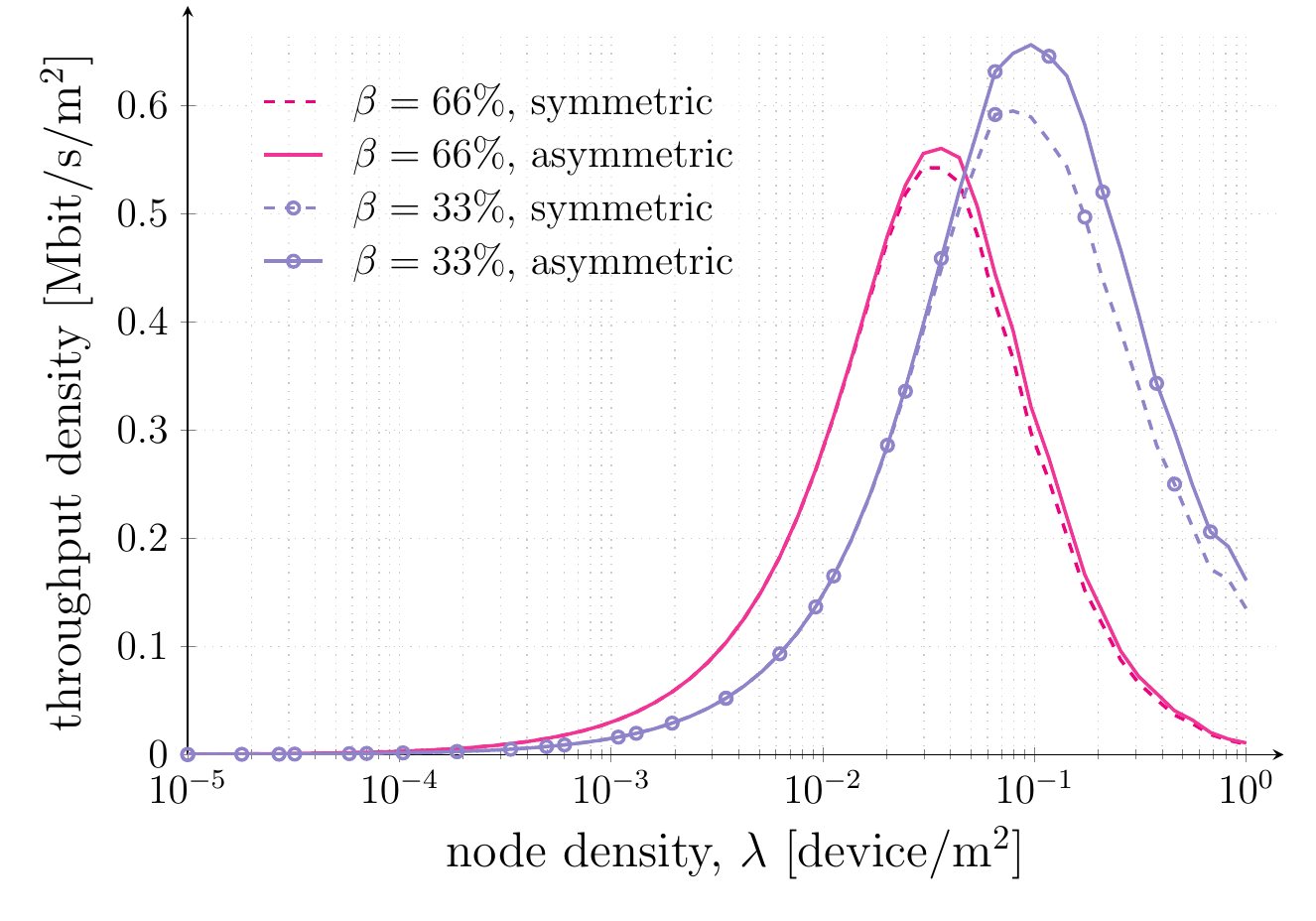}
\vspace{-4mm}
\caption{Throughput density vs. total network density, with different fractions $ \beta $ of communication devices (simulation results). Packet duration $ \pkL=M/2 $ slots, $ \varphi_c = 30^\circ $, and $ \varphi_r = \{30^\circ, 60^\circ\}$ in symmetric and asymmetric antenna configurations, respectively.}
\label{fig:t_60_30}
\end{figure}

Lastly, we study the effect of antenna directionality on the coexistence performance by considering an asymmetric setup of practical engineering interest, where the radars are equipped with less directional antennas than the communication devices. The targets that the radars are attempting to detect can be located in any direction and therefore the wider beam of the antenna can be advantageous for the detection. By contrast, a communication device transmits to a dedicated receiver and can benefit more from the increased gains of a higher directionality. To this end, in Figs.~\ref{fig:dm_60_30}~and~\ref{fig:t_60_30} we present simulation results where radars are utilising an antenna with a wider 60$^\circ$ \ac{HPBW}, while communication devices are still using the reference antenna with a 30$^\circ$ \ac{HPBW}.

Fig.~\ref{fig:dm_60_30} shows that antenna directionality has a significant effect on radar performance. It is evident that, although the larger beamwidth enables detection over a wider field of view, the detectable range is severely reduced. This is a consequence of the increased amount of interference which is received through the wider beam, as well as the lower gain for the detection of the target.  By contrast, Fig.~\ref{fig:t_60_30} shows that the considered asymmetric setup is beneficial in terms of aggregate throughput. Although the communication receivers are now located within the main beam of more radars, the transmit antenna gain of each interferer is lower. Therefore, the overall amount of interference from the radars is reduced and an increase in throughput can be achieved. Moreover, we can observe a bigger increase in throughput when the percentage of radars in the network is higher, confirming that the reduced interference from the rest of the network improves the performance of the communication devices.
Our results thus underline that radar antenna directionality offers an important design trade-off. 

\section{Conclusions}\label{sec:concl}
In this paper, we presented an analysis of the performance of radar and communication networks which are coexisting in the same spectrum band. Considering pulsed radars that follow a regular transmission pattern and communication devices employing a $ p_t $ persistent ALOHA policy, we studied the effect of mutual interference on both systems in different scenarios by means of both a mathematical model for radar performance and of extensive Monte Carlo simulations.  Our study showed that radar detection is severely limited by interference and that deployment in dense topologies envisioned for emerging applications will be challenging and require interference-mitigation techniques. Communication node performance, on the other hand, is not significantly affected by radar operation due to the low transmission activity of the latter. An encouraging result for enabling coexistence is that both systems have transmission patterns which can be beneficial to the other and provide better performance than single-system deployments in networks of equal density. Longer communication packets were also found to positively increase the performance of radars, while the directionality of the antennas was shown to have a significant effect on the detectable range. Our ongoing work is considering different types of radar operations and more sophisticated medium access control protocols for both the communication and radar networks (e.g. CSMA). The study of mobility and radar target tracking performance is also of interest for future work.

\appendix

We aim to compute the number $\omega(\nu_1)$ of transmissions opportunities a communication device \si\ has that overlap with a single echo waiting period  \mbox{$\{nM, \dots, (n+1)M - 1 \}$}, $n\in N$ at the typical radar. By the definition of $\nu_1$, at slot $nM + \nu_1$ node \si\ decides whether to access the channel or defer activity for the subsequent $L$ slots. Therefore, $\omega(\nu_1)$ can be expressed as $1 + \omega_a(\nu_1) + \omega_b(\nu_1)$, where $\omega_a(\nu_1)$ and $\omega_b(\nu_1)$ denote the number of packets that fit -- i.e. at least partially overlap -- within the periods between slots $nM+1$ and $nM+\nu_1-1$ and between slots $nM+\nu_1 + L + 1$ and $(n+1)M-1$, respectively.\footnote{We recall that a transmission over slot $nM$, concurrent with a radar pulse emission, would not cause interference and is not accounted for in $\omega(\nu_1)$.} Recalling that $\nu_1 \in \{ 0, \dots, M-1\}$, we get
\begin{align*}
\omega_a(\nu_1) &= \max\{ 0, \lceil (\nu_1)/L \rceil \} \\
\omega_b(\nu_1) &= \lceil  (M-1 - \min\{ \nu_1 + L- 1, M-1 \} )/L \rceil,
\end{align*}
leading to \eqref{eq:numPkts}. 
\pagebreak
\bibliographystyle{IEEEtran}
\bibliography{IEEEabrv,biblioRadar}

\end{document}